\documentclass[onecolumn,amsmath,amssymb]{qm2008_abs}
\usepackage{graphicx}
\usepackage{epsfig}
\usepackage{dcolumn}
\usepackage{bm}
\begin{document}

\title{{Z$^{0}$ Boson Measurement with the ALICE Central Barrel\\ in $pp$ collisions at 14\,TeV}}

\bigskip
\bigskip
\author{\large R. Bailhache$^1$}
\author{\large A. Andronic$^1$}
\author{\large P.Braun-Munzinger$^{1,2}$ \\ for the ALICE Collaboration}


\affiliation{$^1$Gesellschaft fuer Schwerionenforschung mbH, Darmstadt, Germany}
\affiliation{$^2$Technische Universitaet Darmstadt, Germany}
\bigskip
\bigskip

\begin{abstract}
\leftskip1.0cm
\rightskip1.0cm
The possibility to detect the $Z^{0}$ in the ALICE central barrel is studied via the electronic decay channel $Z^{0}$$\to$$e^{+}e^{-}$. The signal and the background are simulated with the leading order event generator PYTHIA 6. The total cross-sections are taken from NLO calculations. Based on test beam data, the electron identification performance of the Transition Radiation Detector is extrapolated to high momenta. The expected yields for minimum-bias $pp$ collisions at 14\,TeV are presented. An isolation cut on the single electron, together with a minimum transverse momentum cut, allows to obtain a clear signal. The expected background is of the order of 1\,$\%$ with the main contribution coming from misidentified pions from jets.
\end{abstract}

\maketitle

\paragraph*{}

The measurements of the $W^{\pm}$ and $Z^{0}$ bosons in $p\bar{p}$ and $e^{+}e^{-}$ collisions have allowed precise test of the Standard Model (SM) of particle physics. In $pp$ collisions at the LHC, the convergence of the NLO and NNLO calculations offers the possibility to use the total $Z^{0}$ cross section for a better understanding of the collider luminosity and the acceptance and efficiency of the detectors \cite{1}. The high $p_{T}$ electrons emitted in the electronic $Z^{0}$ decays can be a controlled observable for checks of  the $p_{T}$ calibration and resolution between 30\,GeV$/$c and 50\,GeV$/$c. In heavy ion collisions $Z^{0}$ is a good candidate for an alternative reference for quarkonium study, despite the large mass differences, $m_{Z}$$\gg$$m_{J/\Psi}$, and the difference in production mechanisms, mainly $q\bar{q}$ for $Z^{0}$ and $gg$ for quarkonium. It should be weakly affected by nuclear shadowing \cite{2} and the presence of the Quark Gluon Plasma \cite{3}. In this work, a feasibility study is presented to detect $Z^{0}$ through $Z^{0}$$\to$$e^{+}e^{-}$ in the central barrel of ALICE. The detection of the $W^{\pm}$ and $Z^{0}$ bosons through their muon decays in the ALICE muon spectrometer has been previously extensively studied \cite{4}.

\paragraph*{}

The leading order event generator PYTHIA 6.326 \cite{5} is used to simulate the production of $Z^{0}$ bosons. Only the lowest order Born processes, $q\bar{q}$$\to$$\gamma^{*}/Z^{0}$, have been generated. The parton shower algorithm of PYTHIA produces additional jets, that mimic the contributions of higher processes,  $q$$\overline{q}$$\to$$\gamma^{*}/Z$$g$ and $q(\overline{q})$$g$$\to$$\gamma^{*}/Z$$q(\overline{q})$. The CTEQ5L PDFs are used. It was shown that the $p_{T}$ and $y$ $Z^{0}$ distributions measured at Tevatron energies are well reproduced \cite{0}. Pure $Z^{0}$ production, without the complete $\gamma^{*}/Z^{0}$ interference, has been simulated in this work. Due to the large vector boson masses, the contributions of higher order QCD processes can be approximated by a $k$ factor, found to be about 1.5 from comparison with measurements in $p\bar{p}$ collisions. The extrapolated cross sections for the LHC are summarized in Tab.\ref{tableui}. The yields were calculated taking an inelastic $pp$ cross section of 70\,mb at 14\,TeV. Calculations have been carried out up to NNLO. In the following we normalise all the cross section to the NNLO calculations \cite{1}. 

\begin{table}[h]
  \begin{center}
    \begin{tabular} {|c|c|c|c|}
      \hline  $pp$ at 14\,TeV & $\sigma_{PYTHIA}$ [nb] & $\sigma_{NNLO}$ [nb] & $N^{X_{pp}}$  \\\hline
      $Z^{0}$$\to$$e^{+}e^{-}$ & $\approx$2.4 & $\approx$1.84 & 3$\times$$10^{-8}$  \\\hline
      $W^{\pm}$$\to$$e\nu$ & $\approx$23.8 & $\approx$19.8 & 3$\times$$10^{-7}$  \\\hline
    \end{tabular}
    \caption{Inclusive cross sections times branching ratio obtained with PYTHIA and extrapolated after comparison to S$p\bar{p}$S and Tevatron data, leading to a $k$ factor of 1.5. Results are for $pp$ collisions at 14\,TeV and are compared with NNLO calculations \cite{1}.}
    \label{tableui}
  \end{center}
\end{table}
\paragraph*{}

\begin{figure}[h]
  \begin{center}
    \begin{tabular}{cc}
      \begin{minipage}{.48\textwidth}
	\includegraphics[width=1.17\textwidth,height=1.05\textwidth]{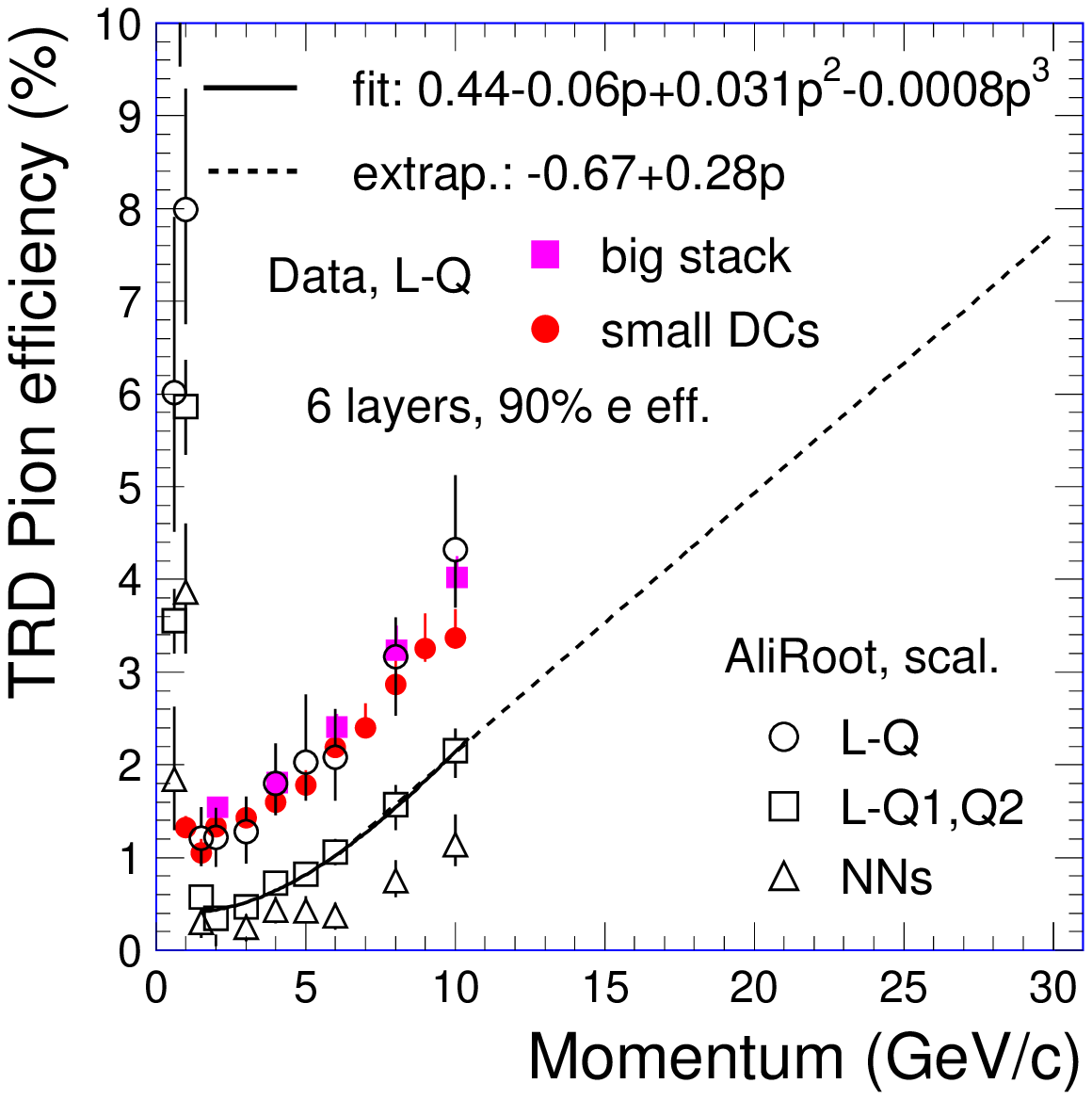}
      \end{minipage} & \begin{minipage}{.48\textwidth}
	\includegraphics[width=1.10\textwidth,height=0.945\textwidth]{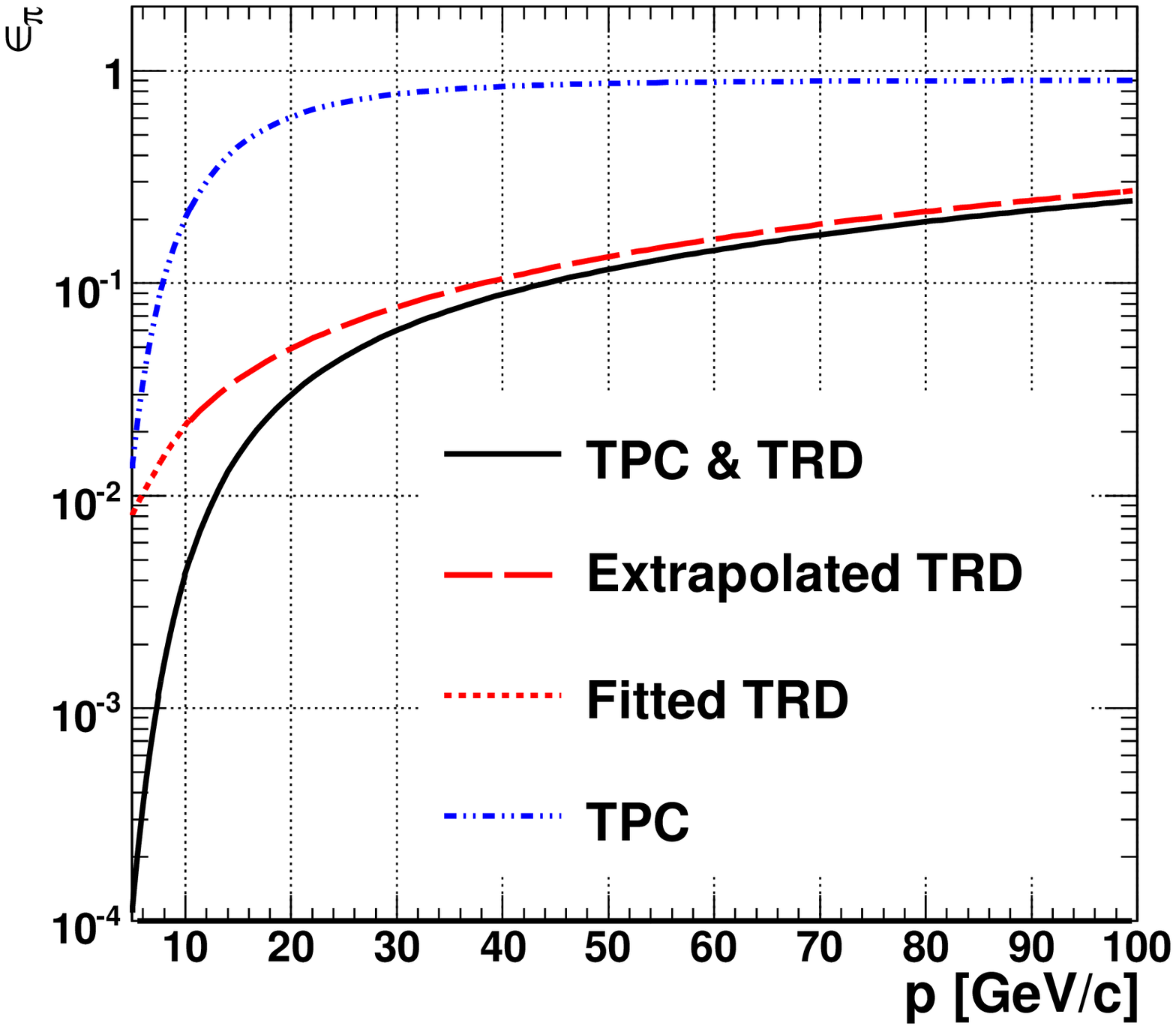}
    \end{minipage} \end{tabular}
  \end{center} 
  \caption{\label{fig-ladutr}$\epsilon_{\pi}^{TRD}$ as function of momentum extracted from testbeam data compared with simulations within the AliRoot framework and extrapolated to high $p$ (left panel). Calculated $\epsilon_{\pi}^{TPC}$, extrapolated $\epsilon_{\pi}^{TRD}$ and combined $\epsilon_{\pi}$ as function of momentum (right panel).}     
\end{figure}
The Inner Tracking System (ITS), Time Projection Chamber (TPC) and Transition Radiator Detector (TRD) provide good tracking capability within their geometrical acceptance, $|\eta|$$<$0.9, 0$<$$\phi$$<$2$\pi$. The Particle Identification (PID) algorithm used requires that the particles are reconstructed in at least five planes of the TRD, which leads to an overall mean reconstruction efficiency of 80\,$\%$. The $p_{T}$ resolution is about 3.5\,$\%$ at 100\,GeV$/$c in the nominal 0.5\,T magnetic field. To identify the electrons, the $dE/dx$ of the TPC and the TRD are used. At such high $p_{T}$, the main difficulty comes from the much more numerous $\pi^{\pm}$ that can be misidentified as electrons. The percentage of misidentified $\pi^{\pm}$, the $\pi^{\pm}$ efficiency $\epsilon_{\pi}$, is determined for a given $e^{\pm}$ efficiency, $\epsilon_{e}$. The left panel of Fig.\ref{fig-ladutr} shows $\epsilon^{TRD}_{\pi^{\pm}}$, as it has been obtained from test beam data analysis of small and big chambers \cite{6} and from simulations done within the AliRoot framework \cite{7}. The results of a one dimensional likelihood method, L-Q, can be improved by using a two dimensional method, L-Q1,Q2 or a neural network, NNs \cite{6}. A fit of the L-Q1,Q2 performances allows to extrapolate $\epsilon_{\pi}^{TRD}$ to the $p$ range of interest for the $Z^{0}$. On the right panel of Fig.\ref{fig-ladutr}, $\epsilon_{\pi}^{TPC}$ has been estimated with simulations for $\epsilon_{e}^{TPC}$=90\,$\%$. The final combined $\epsilon_{\pi}$ for $\epsilon_{e}$=81\,$\%$ (=$\epsilon_{e}^{TPC}$$\times$$\epsilon_{e}^{TRD}$=0.9$\times$0.9\,) is also plotted in Fig.\ref{fig-ladutr}. The response of the ALICE central barrel is simulated with a fast simulation program.\\

\begin{figure}[h]
  \begin{center}
    \begin{tabular}{cc}
      \begin{minipage}{.45\textwidth}
	\includegraphics[width=1.25\textwidth,height=1.02\textwidth]{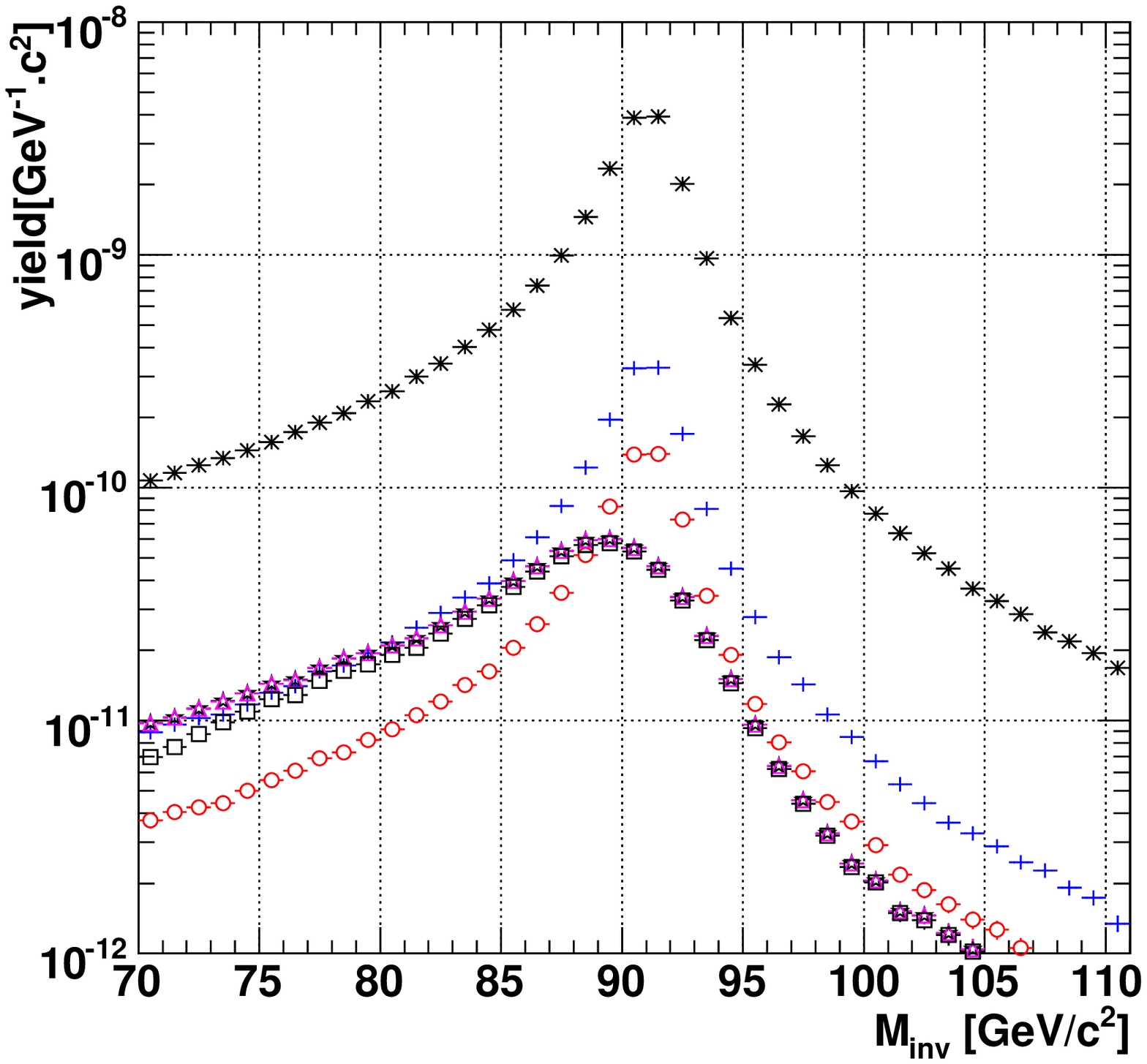}
      \end{minipage} & \begin{minipage}{.45\textwidth}
	\includegraphics[width=0.85\textwidth,height=0.92\textwidth]{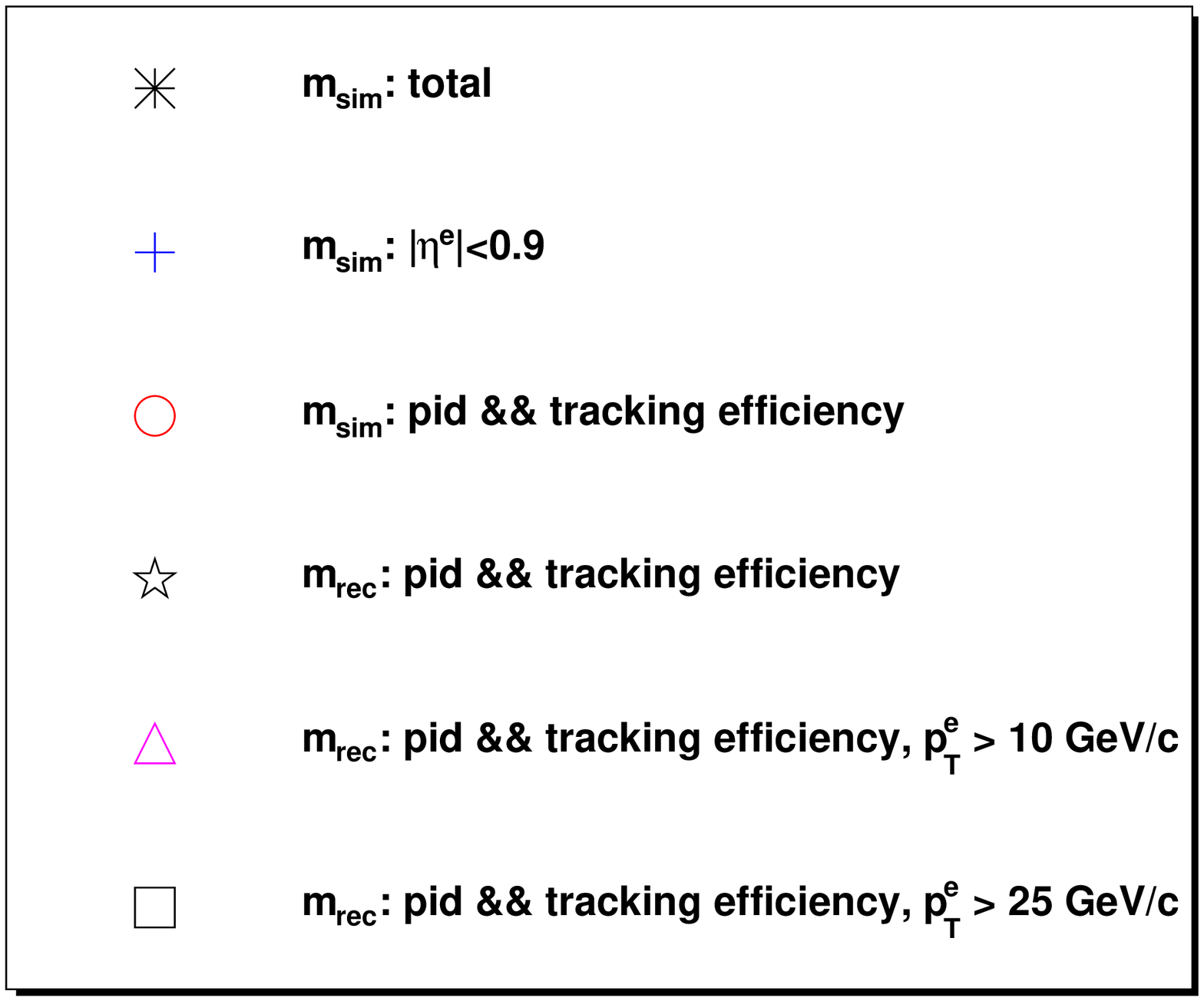}
    \end{minipage} \end{tabular}
  \end{center} 
  \caption{\label{fig-ladutree}$e^{+}e^{-}$ generated ($m_{sim}$) and reconstructed ($m_{rec}$) invariant mass yield from $Z^{0}$ in the total phase space and within the central barrel acceptance for different $p_{T e^{\pm}}$ cut.}     
\end{figure}

\begin{table}[h]
  \begin{center}
    \begin{tabular} {|c|c|}
      \hline
      \,$|\eta_{e^{+/-}}|$$<$0.9\, & \,8.6\,$\%$\,\\\hline
      \,${\it{A_{e}}}$$\times$$\epsilon^{tr}_{e}$$\times$$\epsilon^{pid}_{e}$\,& \,3.5\,$\%$\,\\\hline
      \,${\it{A_{e}}}$$\times$$\epsilon^{tr}_{e}$$\times$$\epsilon^{pid}_{e}$ $p_{T e^{\pm}}$$>$25\,GeV$/$c\,& \,3.2\,$\%$\,\\\hline
      \,${\it{A_{e}}}$$\times$$\epsilon^{tr}_{e}$$\times$$\epsilon^{pid}_{e}$ $p_{T e^{\pm}}$$>$25\,GeV$/$c iso cut\,& \,3.2\,$\%$\,\\\hline
    \end{tabular}
  \end{center}
  \caption{Acceptance and reconstruction efficiency for $Z^{0}$ in the mass range 60\,GeV$/$c$^{2}$$<$M$_{e^{+}e^{-}}$$<$116\,GeV$/$c$^{2}$ for different single track cuts.}
  \label{opi}
\end{table}

The geometrical acceptance of the central barrel implies that both of the electrons have $|\eta_{e^{\pm}}|$$<$0.9. This reduces the $Z^{0}$ yield to 8.6\,$\%$ of the full phase space yield (see Tab.\ref{opi}). The statistical errors are below 1\,$\%$. A clear signature of $Z^{0}$ decays is two high $p_{T}$ isolated electrons. A $p_{T}$ cut at 25\,GeV$/$c is considered together with an isolation cut. It will reject a track $i$, if a track $j$ is found to have: $p_{T}^{j}$$>$2\,GeV, $|\eta_{i}-\eta_{j}|$$\leq$0.1 and $|\phi_{i}-\phi_{j}|$$\leq$0.1\,rad. 99\,$\%$ of the signal survives this cut. Fig.\ref{fig-ladutree} shows the generated $m_{sim}^{e^{+}e^{-}}$ in the total phase space and in the geometrical acceptance with and without tracking and PID efficiencies. The reconstructed $m_{rec}^{e^{+}e^{-}}$ is also plotted for different $p_{T e^{\pm}}$ cuts. Bremsstrahlung leads to a tail towards lower values of the mass.

\paragraph*{}

The different sources of background that are investigated in $pp$ collisions at $\sqrt{s}$=14\,TeV are: reconstructed dielectrons from jets, that can be real electrons or pions misidentified as electrons; $W$$\to$$e\nu$ events with an associated hadronic jet that results in a second reconstructed electron ($Br_{W\to e\nu}$=10.75\,$\%$); $Z^{0}$$\to$$\tau\tau$ events, in which electrons or misidentified pions from $\tau$ decays ($Br_{\tau \to e/\pi+X}$=44.0850\,$\%$) are combined; electrons and misidentified pions from $t\bar{t}$ events ($Br_{t\to bW}$$\approx$100\,$\%$); simultaneous semielectronic decays of $D$ and $\bar{D}$ mesons (Br$_{c\to eX}$$\approx$9.6\,$\%$); and simultaneous semielectronic decays of $B$ and $\bar{B}$ mesons (Br$_{b\to eX}$$\approx$10.86\,$\%$).
The jets have been simulated with the PYTHIA using Tune A CDF, that gives a total cross section of 54.7\,mb \cite{8}. Due to the high masses of the $W$ boson and the top quark, only the lowest order processes for $W$ production ($q$$\bar{q^{'}}$$\to$$W$) and $t\bar{t}$ production ($g$$g$$\to$$t\bar{t}$ and $q$$\bar{q}$$\to$$t\bar{t}$) have been generated with PYTHIA and normalised to the NLO cross sections. For the lighter $c$ and $b$ quarks production, contributions from higher order corrections, like flavour excitations ($q$$Q$$\to$$q$$Q$) and gluon splitting ($g$$\to$$Q\bar{Q}$) have also been taken into account. The tuned PYTHIA \cite{10} $p_{T}$ spectra of $c$ and $b$ have been compared to NLO predictions (HVQMNR program \cite{9}) and found to be softer by an order 10 at very high $p_{T}$. This would result in a contribution of $c\bar{c}$ and $b\bar{b}$ about 100 higher in the invariant mass yield.\\
 \begin{figure}
   \begin{center}
     \begin{tabular}{cc}
       \begin{minipage}{.48\textwidth}
	 \includegraphics[width=1.15\textwidth,height=0.96\textwidth]{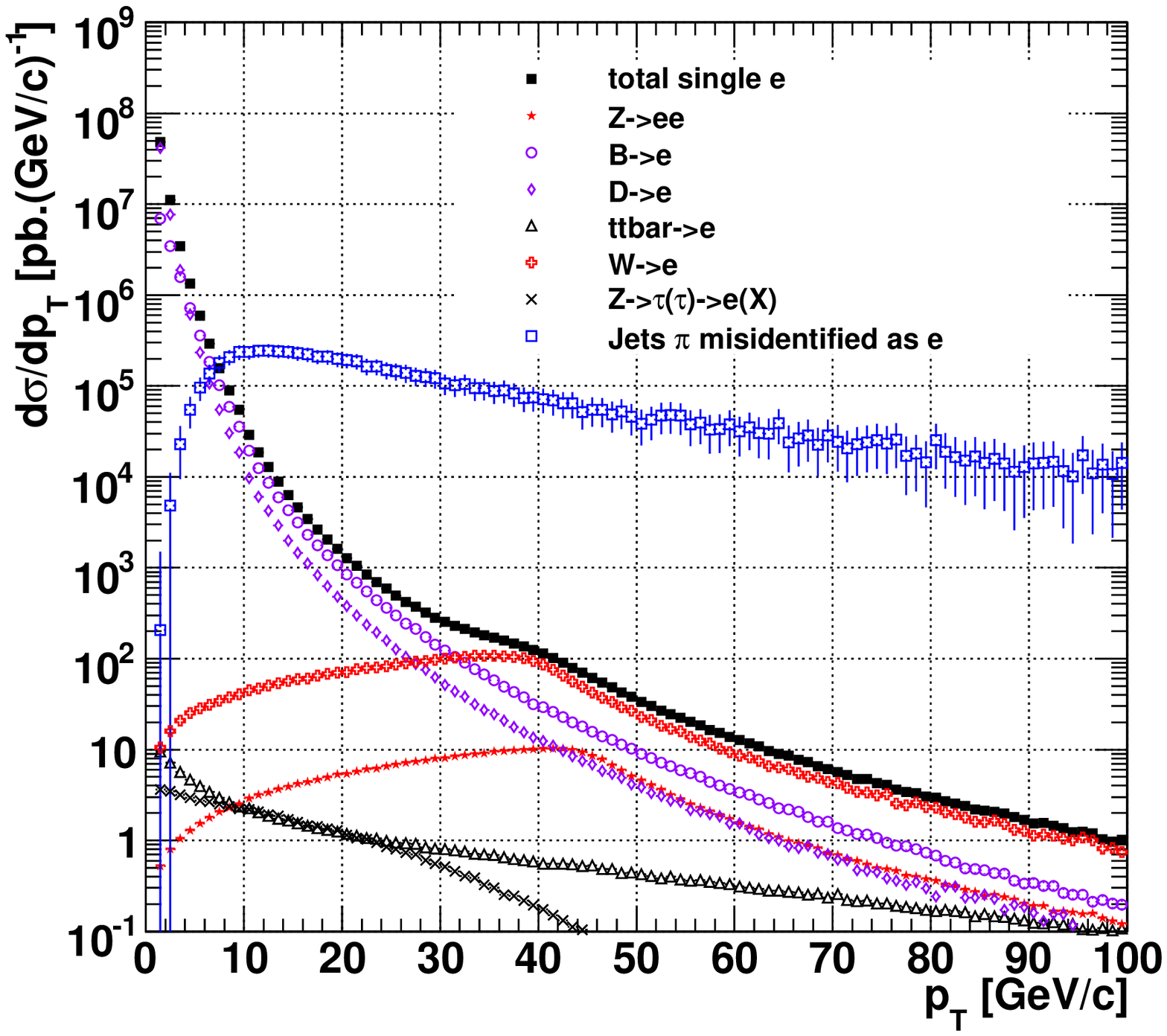}
       \end{minipage} & \begin{minipage}{.48\textwidth}
	 \includegraphics[width=1.15\textwidth,height=0.96\textwidth]{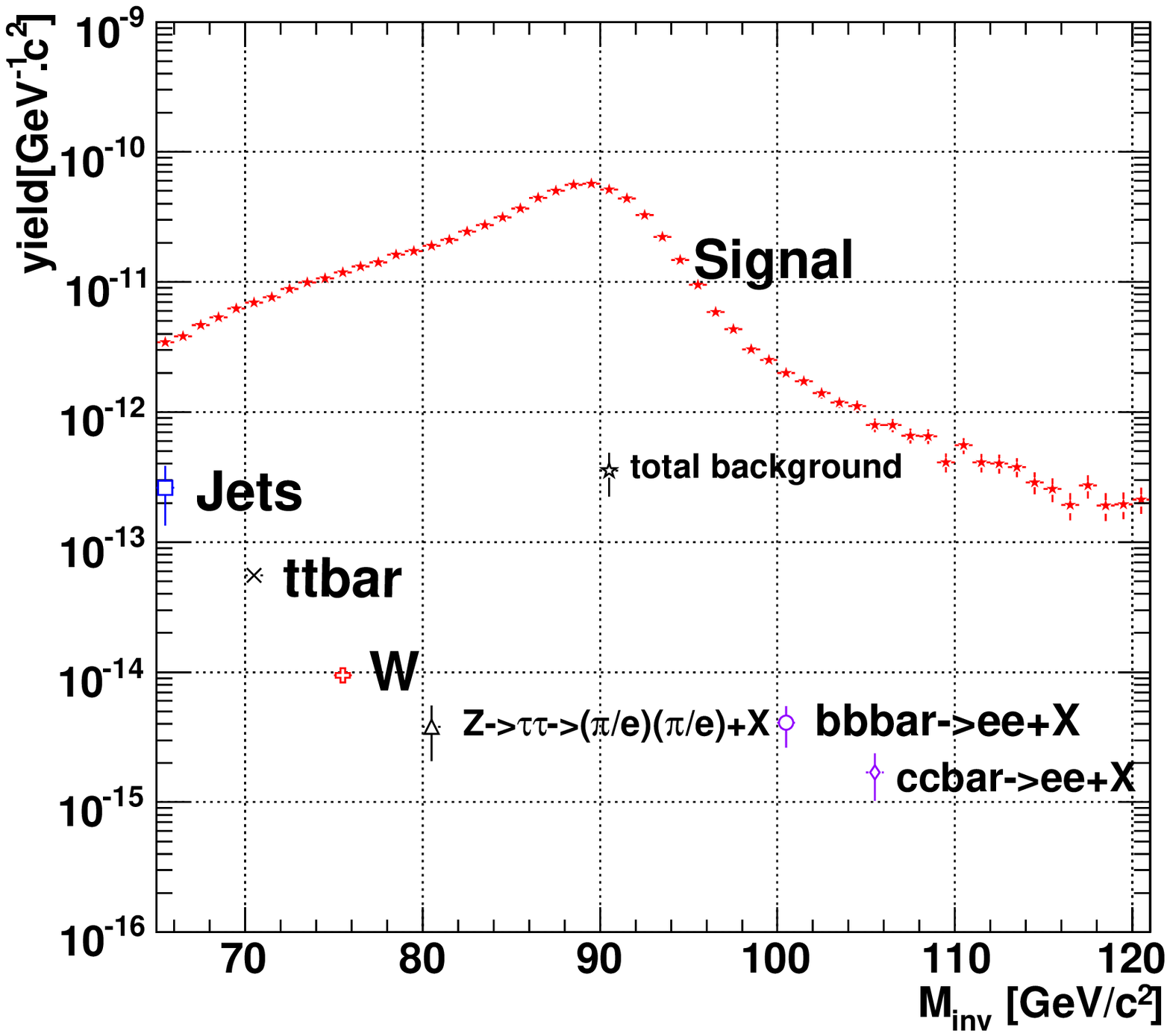}
     \end{minipage} \end{tabular} 
   \end{center}
   \caption{\label{fig-ouilop} Left panel: reconstructed single electron spectra in the central barrel. Right panel: comparison with Z$^{0}$ signal of different contributions to the background for a $p_{T}$ cut at 25\,GeV$/$c and the isolation cut. The contributions have been averaged over the invariant mass range 66\,GeV$/$c$^{2}$$<$$M_{e^{+}e^{-}}$$<$116\,GeV$/$c$^{2}$.}     
\end{figure} 
The left panel of Fig.\ref{fig-ouilop} shows the reconstructed electron spectra. Misidentified $\pi^{\pm}$ from jets constitute the main source of reconstructed electrons above 10\,GeV$/$c. Nevertheless they are not isolated. The rejection factor of the isolation cut is of the order of 10$^{4}$. The different contributions to the dielectron reconstructed invariant mass yield per minimum-bias $pp$ collisions are presented in the right panel of Fig.\ref{fig-ouilop}. A $p_{T}$ cut at 25\,GeV$/$c and the isolation cut are applied. The isolation cut suppresses also the correlated background from simultaneous semi-electronic decays of $D$ and $\bar{D}$, or $B$ and $\bar{B}$, mesons, below one percent, even with a factor 100, due to higher order corrections. The final total background amounts to about (0.7$\pm$5.3)\,$\%$ of the signal, with a main contribution from misidentified pions from jets. The errors given are statistical.

\paragraph*{}

We have presented a study of $Z^{0}$ reconstruction in $pp$ collisions at 14\,TeV with the central barrel of the ALICE detector. The $Z^{0}$$\to$$e^{+}e^{-}$ yields are of the order of 3$\times$10$^{-8}$ per minimum-bias $pp$ collisions. A Level 1 TRD trigger (p$_{T}$$>$10\,GeV$/$c) for 10\,$\%$ of data taking time would lead to a $Z^{0}$ sample of about 100 per year. Further enhancement is possible using the High-Level Trigger. The decay electrons are identified with the TRD and the TPC detectors within the central barrel ($|\eta|$$<$0.9). The probability to misidentify a $\pi^{\pm}$ has been extrapolated to the high momentum region of interest and is of the order of 0.1 at 45\,GeV$/$c. The two main characteristics of the electrons emitted in $Z^{0}$ decays, i.e. high $p_{T}$ and isolation, have been used to reject the background. Two high $p_{T}$ isolated reconstructed electrons constitute a very clear signature of the $Z^{0}$ in the central barrel. The background is expected to be of the order of 1\,$\%$ in $pp$ collisions, dominated by misidentified pions from jets.\\
{\bf{Acknowledgment:}}\\
We thank Chuncheng Xu for pointing out the importance of the isolation cut.


\noindent
\begin{thebibliography}{50}
\bibitem{1} A.D. Martin et al., Eur. Phys. J.C. 18, 117-126 (2000)
\bibitem{2} R. Vogt, Phys. Rev. C 64, 044901 (2001)
\bibitem{3} J.I. Kapusta and S.M.H. Wong, Phys. Rev. D 62, 037301 (2000)
\bibitem{4} Z. Conesa del Valle et al., Phys. Lett. B 663, 202 (2008), arXiv:hep-ph/0702118v2
\bibitem{5} T. Sjostrand et al., PYTHIA 6.3 Physics and Manual, hep-ph/0308153 (2003)
\bibitem{0} Britta Tiller for the D0 Collaboration, DPG Berlin 04/03/2005; R.Bailhache to be published 
\bibitem{6} A. Andronic et al., Nucl. Instr. and Meth. A 522 40 (2004); ALICE TRD Collaboration, GSI Scientific Report (2004) 355; A. Wilk et al., Nucl. Instr. Meth. A 563 310 (2006)
\bibitem{7} AliRoot, An Object-Oriented Data Analysis Framework, http://aliweb.cern.ch/offline/
\bibitem{8} R. Field http://www.phys.ufl.edu/$\sim$rfield/cdf/tunes/py$_{-}$tuneA.html
\bibitem{10} Alice Physics Performance Report Volume II, J.Phys. G 32 1753-1843 (2006)
\bibitem{9}  M. Mangano, P. Nason and G. Ridolfi, Nucl. Phys. B 373 295 (1992) 
\end{thebibliography}
\end{document}